\documentclass{phb-proc4-auth}

\usepackage{graphicx}
\usepackage{amssymb,amsmath}

\begin{document}
\begin{frontmatter}


\journal{SCES '04}


\title{ Melting of the stripe phases in the $t$-$t'$-$U$ Hubbard model }

%

\author [FR,PL]{Marcin Raczkowski,\thanksref{KBN}}
\author[FR]{Raymond Fr\'esard,\corauthref{cor}}
\corauth[cor]{Corresponding author (Raymond Fr\'esard).\newline 
{\it Email address:\/} Raymond.Fresard@ensicaen.fr}
\author[PL]{Andrzej M. Ole\'s}

%

\address[FR]{Laboratoire CRISMAT, UMR CNRS--ENSICAEN(ISMRA) 6508,
6 Bld. du Mar\'echal Juin, F-14050 Caen, France}
\address[PL]{Marian Smoluchowski Institute of Physics, Jagellonian
 University, Reymonta 4, PL-30059 Krak\'ow, Poland}


\thanks[KBN]{M.R. was supported by a Marie Curie fellowship of the 
             European Community program under number HPMT2000--141.
             This work was supported by the Polish State Committee of 
             Scientific Research (KBN) under Project No. 1~P03B~068~26.}

%
%
%
%

\begin{abstract}
We investigate melting of stripe phases in the overdoped regime 
$x\geqslant 0.3$ of the two-dimensional $t$-$t'$-$U$ Hubbard model, 
using a spin rotation invariant form of the slave boson representation.
We show that the spin and charge order disappear simultaneously, and 
discuss a mechanism stabilizing bond-centered and site-centered stripe 
structures.
\end{abstract}

%



\end{frontmatter}

%

It is now well established that the doped cuprates show many highly 
unusual properties both in normal and superconducting state. Among 
them, stripe phase, discovered in theory \cite{Zaa89} and confirmed 
by experiment \cite{Tran95}, attracted a lot of interest. 
Instead of moving independently, the holes introduced to an 
antiferromagnetic (AF) Mott insulator self-organize either on 
site-centered (SC) nonmagnetic domain walls (DW) separating AF
spin domains, or on bond-centered (BC) DW made out of pairs of 
ferromagnetic spins \cite{Tran95}. Such a tendency towards phase 
separation is fascinating, and offers a framework for interpreting 
a broad class of experiments, including the pseudogap at the Fermi
energy observed in the angle-resolved photoemission (ARPES) spectra 
of La$_{2-x}$Sr$_x$CuO$_4$ (LSCO) for the entire underdoped regime 
($0.05\leqslant x\leqslant 0.125$), reproduced within the $t$-$t'$-$U$ 
Hubbard model \cite{Fleck01}. Therefore, we argue that this model is 
sufficient to investigate generic features of stripe phases.

Two main scenarios for a driving mechanism of the stripe phase have 
been proposed \cite{Zach98}. In the first one stripes arise from a 
Fermi surface instability with the spin driven transition \cite{Zaa89}; 
then spin and charge order simultaneously, or charge order follows spin 
order. The second scenario comes from Coulomb-frustrated phase
separation suggesting that stripe formation is commonly charge driven, 
and the charge order sets in first when the temperature is lowered. 
However, slave boson studies of the two-dimensional (2D) $t$-$t'$-$U$ 
Hubbard model showed that the spin susceptibility diverges while the 
charge susceptibility does not \cite{Fre98}, so the microscopic origin 
of the stripe instability is unclear.

We investigate the mechanism leading to phase separation and the 
melting of vertical BC and SC stripe phases in the overdoped regime 
($x\geqslant 0.3$, where $x=1-n$ and $n$ is an average electron density
per site) of the 2D $t$-$t'$-$U$ Hubbard model. We employ the spin 
rotation invariant slave boson (SB) representation of the Hubbard model 
\cite{Fre92}, and perform the calculations on larger (up to 
144$\times144$) clusters than those studied recently \cite{Sei04}. 
This allows one to obtain unbiased results at low temperature $T=0.01t$.
\begin{figure}[t!]
\begin{center}
\includegraphics[scale=0.42]{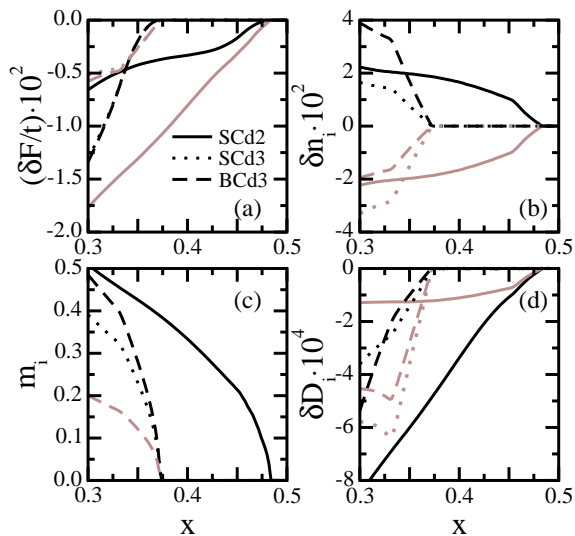}
\end{center}
\caption
{
Melting of vertical BC and SC with increasing doping $x$ at $T=0.01t$: 
(a) the free energy $\delta F$ (black line) and interaction (grey line) 
energy gain in the stripe phases; 
(b) local charge densities $\delta n_i$ relative to their average 
values;
(c) local magnetization $m_i$;
(d) double occupancies $\delta D_i$ relative to the values in the 
paramagnetic phase (scaled by a factor $\tfrac{1}{3}$ for the $d=2$ 
stripe). In panels (b)-(d) the black (grey) curves correspond to the 
strongly (weakly) polarized sites, respectively.
} 
\end{figure}
 
For the model parameters for LSCO: $U/t=12$ and $t'/t=-0.15$,
we obtain that the most stable SC stripes are separated by $d=4$(3) 
lattice spacings at dopings $0.124\leqslant x\leqslant 0.2$ 
($0.2\leqslant x\leqslant 0.34)$, respectively. As shown in Fig.~1(a), 
increasing doping stabilizes the SC stripes with a single atom in the 
AF domains. Also for the BC stripes the size of the AF domains 
decreases with increasing doping, varying from $d=5$
($0.10\leqslant x\leqslant 0.13)$ through $d=4$ 
($0.13\leqslant x\leqslant 0.19)$ and down to $d=3$ at higher doping, 
as there is no BC configuration with $d=2$. For both types of stripes, 
the distance between them is locked to four in a sizeable doping range 
above $x\simeq \tfrac{1}{8}$, in agreement with neutron scattering 
experiment \cite{Yam98} and with theory \cite{Fleck01} for LSCO. 

In Fig.~1(a) we show the energy gain of the stripe phases with respect 
to the paramagnetic phase $\delta F$. Remarkably, the difference in 
energy between the best SC and BC stripes is smaller than both the 
accuracy of the calculations, and the resolution of Fig.~1(a), 
suggesting that quantum fluctuations might be important.
We characterize the melting of stripes by their SB local averages: 
density $n_i=\sum_{\sigma}\langle n_{i\sigma}\rangle$, magnetization 
$m_i=|\langle S_{i}^z\rangle|$, and double occupancies
$D_i=\langle n_{i\uparrow}n_{i\downarrow}\rangle$. 

In the $d=2$ SC 
stripe, reported here for the first time, the two $\delta n_i(x)$ curves 
are symmetrical in Fig.~1(b). In contrast, in the $d=3$ BC stripe there 
are two sites with weak magnetic moments per one strongly polarized 
site. We note that, unlike in the SC phase, the variation in density is
largest on the strongly polarized sites in the BC phase. 
The magnetic moments $m_i$ vanish for both types $d=3$ stripes at the same 
doping $x=0.375$ [Fig.~1(c)], suggesting that they originate from the same 
instability. 

The microscopic mechanism stabilizing the $d=2$ SC stripes appears 
to differ markedly from the one stabilizing the $d=3$ ones \cite{Oles}. 
For $d=2$ [Fig.~1(d)], the reduction of double occupancy is strongest on the
magnetic sites, and the corresponding reduction
of interaction energy is larger than the gain of free energy 
[see Fig.~1(a)]. Thus the mechanism leading to the formation of the 
$d=2$ stripe is primarily local, making use of two complementary 
effects helping to reduce double occupancy: finite magnetization at 
magnetic sites and reduced electron density at nonmagnetic ones.
Even though such a state looses kinetic energy, the gain in the 
interaction energy overcompensates this loss, stabilizing this order 
in a wide doping range $x\leqslant 0.485$.

In contrast, for $d=3$ stripes, both contributions to the free energy
are substantially decreased while stripe order starts melting 
already at $x<0.3$ mainly by faster removing double occupancies from 
the stripe DW than from AF domains leading to gradually disapearing 
magnetic moment upon doping. Therefore, both potential and kinetic 
energy (including the superexchange) cooperate to stabilize stripes 
with $d>2$. In fact, for both $d=3$ stripes, the mechanism is doping 
dependent. In the small magnetization regime, the interaction energy
plays the leading role.  
However, under a further decrease of hole density, this gain nearly 
saturates (at $x\simeq 0.33$), and the gain in the kinetic energy 
starts to dominate. Moreover, it is only slightly larger for the SC 
stripe compared to the BC one, and therefore it is easily compensated, 
mainly by the presence of finite magnetic moments at BC domain walls. 
As a common feature, the spin and charge order {\it disappear\/} at the 
same critical doping. Therefore, in the absence of longer ranged Coulomb 
interaction the charge order is always accompanied by the spin order. 

Summarizing, we have investigated the microscopic mechanisms
responsible for the formation of the vertical BC and SC stripes in 
the extended 2D Hubbard model. Interestingly, we found that BC and 
SC stripes remain nearly degenerate, and both {\it spin and charge\/} 
order vanish simultaneously when they melt, demonstrating a cooperative 
character of the stripe order.


\end{document}